\documentclass[10pt,twocolumn,aps,prl,showpacs,superscriptaddress,floatfix]{revtex4-1}

\usepackage{graphicx}
\usepackage{amssymb}
\usepackage{amsmath}
\usepackage{color}
\usepackage{psfrag}
\usepackage{epsfig}
\usepackage{bbm}
\usepackage{bm}
\usepackage{hyperref}
\hypersetup{breaklinks}

\newcommand{\ket}[1]{| #1 \rangle}

\newcommand{\rb}[1]{\left( #1 \right)}

\newcommand{\ew}[1]{\langle #1 \rangle}
\newcommand{\beq}{\begin{eqnarray}}
\newcommand{\eeq}{\end{eqnarray}}

\newcommand{\op}[2]{| #1 \rangle \langle #2 |}

\newcommand{\eq}[1]{Eq.~(\ref{#1})}
\newcommand{\fig}[1]{Fig.~\ref{#1}}

\newcommand{\trace}[1]{\mathrm{Tr}\left\{#1\right\}}

\newcommand{\citer}[1]{{Ref.~\cite{#1}}}

\begin{document}
\title{Temporal quantum correlations and Leggett-Garg inequalities in multi-level systems}
\author{Costantino Budroni}
\affiliation{
  {Naturwissenschaftlich-Technische Fakult\"at, Universit{\"a}t Siegen,
Walter-Flex-Str.~3, D-57068 Siegen, Germany}
}
\author{Clive Emary}
\affiliation{
  Department of Physics and Mathematics,  University of Hull,
  Kingston-upon-Hull, United Kingdom
}

\date{\today}
\begin{abstract}
We show that the quantum bound for temporal correlations in a Leggett-Garg test, analogous to the Tsirelson bound for spatial correlations in a Bell test, strongly depends on the number of levels $N$ that can be accessed by the measurement apparatus via projective measurements.  
We provide exact bounds for small $N$, that exceed the known bound for the Leggett-Garg inequality, and show that in the limit $N\rightarrow \infty$ the Leggett-Garg inequality can be violated up to its algebraic maximum.
\end{abstract}
\maketitle

{\it Introduction.---}  Bell inequalities place fundamental bounds on the nature of correlations between spatially-separated entities within a local hidden variable framework \cite{Bell1964}. 
Leggett and Garg showed that {\em temporal} correlations obey similar inequalities based on assumptions of macroscopic realism and non-invasive measureability \cite{Leggett1985}.
Quantum particles are bound neither by local hidden variables nor macroscopic realism and so can violate both Bell and Leggett-Garg inequalities (LGIs).  

The maximum degree to which a quantum system can violate a Bell inequality is known as the Tsirelson bound \cite{Cirelson1980}, significantly less than the largest-conceivable value, the algebraic bound 
\cite{CHSH}. 
Violations of a Bell inequality beyond the Tsirelson bound would be evidence of a new physics beyond quantum theory \cite{Popescu1994}.

With interest in the LGIs growing (see \citer{Emary2013} for a review), we ask here whether there is such a thing as a {\em temporal Tsirelson bound} for the LGIs.
In the light of some recent results \cite{Fritz2010, Budroni2013} and given the formal symmetry between the two types of inequality \cite{Markiewicz2013,Das2013} and the general trend towards unification between temporal and spatial correlations \cite{Brukner2004,Marcovitch2011a,Oreshkov2012,Fitzsimons2013}, one would expect that the Tsirelson bound for the LGIs holds analogously to the spatial case. 
Surprisingly, and as we show here, this is not the case.
By considering a broader class of  projective measurements than hitherto considered, we show that the maximum quantum violation of the LGIs can exceed the Tsirelson value, and increases with increasing system size, even up to the algebraic bound in the asymptotic limit.

Let us now be more concrete and consider the simplest LGI which, for dichotomous observable $Q=\pm1$, reads
\beq
  K_3 \equiv C_{21} + C_{32} - C_{31} \le 1
  \label{K3intro}
  ,
\eeq 
where $C_{\beta \alpha} = \ew{ Q(t_\beta)Q(t_\alpha)}$ is the
correlation function of variable $Q$ at the two times $t_\beta \ge t_\alpha$.
For a two-level system, the maximum quantum value of $K_3$ is $K_3^\mathrm{max} = \frac{3}{2}$ \cite{Leggett1985}, which we shall refer to as the {\it L\"uders bound}, $ K_3^{\text{L\" uders}} =\frac{3}{2}$, for reasons to become clear shortly.   
It has been proven rigorously that for measurements given by just two projectors, $\Pi_+$ and $\Pi_-$, onto eigenspaces associated with results $Q=+1$ and $Q=-1$, the maximum quantum value of $K_3$ is the same as for the qubit, {\em irrespective of system size} \cite{Budroni2013}.   This has been reflected in several studies: the experiment of \citer{George2013} on a three-level system obtained a maximum value less than $\frac{3}{2}$; on the theory side,  multi-level quantum systems such as a large spin \cite{Kofler2007}, optoelectromechanical systems \cite{Lambert2011} and photosynthetic complexes \cite{Wilde2010} have also been observed to obey  $K_3 \le K_3^{\text{L\" uders}}$.
From this, one might conclude that nothing new is to be gained from considering higher dimensional systems.  Were this the case, the bound for the qubit would apply in all generality and $K_3^{\text{L\" uders}}$ could be identified with the relevant temporal Tsirelson bound. However, as we will show, with a more general projective measurement scheme, violations of \eq{K3intro} for multi-level systems can exceed the qubit value. 

Other than in an invasive scenario (where the algebraic maximum is trivially achieved, e.g., a classical device with memory  or its quantum realization via positive-operator valued measures (POVMs) \cite{Fritz2010}), the only hint that a violation of \eq{K3intro} greater than $K_3^{\text{L\" uders}}$ is possible has come in the recent work by Daki\'c~{\it et al.}~\cite{Dakic2013}. There, however, the excess violation was claimed to stem from correlations beyond quantum theory.
In contrast, our excess violations are found within the standard framework of quantum theory and projective measurements.
This we achieve by considering measurements on an $N$-level system that can project the state in one of $M$ different subspaces, $2\le M \le N$,  with outcomes that are nevertheless associated with either $Q=+1$ or $Q=-1$. From a macroscopic-realist point-of-view, this leaves \eq{K3intro} unchanged. From a quantum perspective, however, the choice of $M$ determines the state-update rule under projective measurement: for $M=2$ the projection is onto one of two subspaces, corresponding to L\"uders rule for dichotomic measurements \cite{Lueders1951}; whereas $M=N$ is the case of a complete degeneracy-breaking measurement, as initially proposed by von Neumann \cite{vonNeumann1932} (see also Ref.~\cite{Lueders2006} for a discussion).  These additional possibilities for state reduction are ultimately responsible for the increased violations.
 
In the present paper we use the example of a large spin precessing in a magnetic field to demonstrate that violations $K_3>\frac{3}{2}$ are possible and that the algebraic bound $K_3 =3$ can be reached. 
We then discuss the exact upper bounds for small $M\leq 5$, and how they may be obtained with few-dimensional systems with $N\leq 9$.

We discuss how a similar modification to the spatial Bell scenario does not lead to an increase in the Tsirelson bound for the corresponding Bell inequality \cite{Cirelson1980}.  Our results therefore reveal a stark contrast between spatial and temporal correlations .  Furthermore, these results imply the utility of the LGIs with our  extended measurement scheme as {\em dimension witnesses} \cite{Brunner08}, i.e., a 
certification of the minimum number of quantum levels an experimenter is able to manipulate, or in 
the discrimination of L\"uders and von Neumann state-update rules \cite{Hegerfeldt2012}.

{\it Preliminary notions.---} We consider measurements of a macroscopic property $Q$, which can take values $\pm 1$, on a $N$-level quantum system, with each level associated with a definite value of $Q$. From a macrorealist point-of-view, the fact that different levels are associated with the same value of $Q$ is irrelevant: they may be considered as microscopically distinct states that have the same macroscopic property $Q$. Macrorealism and non-invasive measurability imply that at each instant of time, the system has a definite value of $Q$, which is independent of measurements previously performed on the system and, therefore, that the bound for Eq. (\ref{K3intro}) in macrorealist theories remains the same. 

From a quantum mechanical perspective, the fact that the system has more than two levels, allows for many possible state-update rules. According to L\"uders' rule  \cite{Lueders1951}, the state is updated as $\rho \mapsto \Pi_\pm \rho \Pi_\pm$, up to normalization, depending on the outcome of the measurement. On the opposite side, von Neumann's original proposal \cite{vonNeumann1932} is  a
state-update  $\rho \mapsto \sum_k \Pi_\pm^{(k)} \rho \Pi_\pm^{(k)}$, where
$\Pi_\pm^{(k)}$ are one-dimensional projectors. Both state-update rules are plausible, and the choice of the correct one depends on the particulars of the interaction between the system and the measurement apparatus (see Ref.\cite{Lueders2006} for a discussion). 

More generally, we consider all possible intermediate cases, namely, state-update rules given by $M$ different projectors, with $2\le M\leq N$, associated with either $+1$ or $-1$ outcome. 
The correlation functions are therefore given by
\beq
  C_{\beta \alpha}\label{QQ}
  =
  \sum_{l,m} q_l q_m 
  \trace{
    \Pi_m U_{\beta \alpha} \Pi_l U_{\alpha 0} 
      \rho_0 
    U^\dag_{\alpha 0} \Pi_l U^\dag_{\beta \alpha}
  }
  ,
\eeq
where $q_l$ represent the outcome $\pm 1$ associated with $\Pi_l$,  $\rho_0$ is the initial state 
of the system and ${U_{\beta \alpha} = U(t_\beta - t_\alpha)}=e^{-i H (t_\beta-t_\alpha)}$ is the 
unitary time-evolution operator for some Hamiltonian $H$.

{\it A simple example.---}Consider a quantum-mechanical spin of length $j$ in a magnetic field oriented in the $x$-direction. We write its Hamiltonian ($\hbar = 1$) as 
\beq
  H = \Omega J_x
  ,
\eeq
with $\Omega$ the level spacing and $J_x$ the $x$-component of the angular momentum operator.
Let us choose to measure the spin in the $z$ direction such that the measurement projectors are $\Pi^j_m = \op{m;j}{m;j}$ with $\ket{m;j}$ eigenstates of the $J_z$ operator.  In this example, we only consider the von Neumann limit, $M=N = 2 j+1$, and choose the measurement values to be $q^j_m = 1-2\delta_{m,-j}$, such that the lowest energy state is associated with the value $-1$, and the rest with $+1$.

Calculating the correlation functions $C_{\beta\alpha}$ for this setup, several differences with the qubit case are immediately apparent.
Most importantly, the correlation functions here depend on both times, not just their difference. As corollary, the correlation functions depend on the initial state.  A further difference is that, for the projectively-measured correlation functions discussed here, the order of the measurements $t_\beta > t_\alpha$ is important.  This is not the case for $M=2$, for which Fritz \cite{Fritz2010} has shown that, for arbitrary $N$, the projectively-measured correlation functions are equal to the expectation value of the symmetrised product $\frac{1}{2} \left\{{Q}_j,{Q}_i\right\}$, where the operators $Q$ have spectral decomposition $Q=\Pi_{+} - \Pi_{-}$, with $\Pi_\pm$ the projectors associated with the eigenvalues $\pm 1$.

\begin{figure}[tb]
  \begin{center}
    \includegraphics[width=\columnwidth,clip]{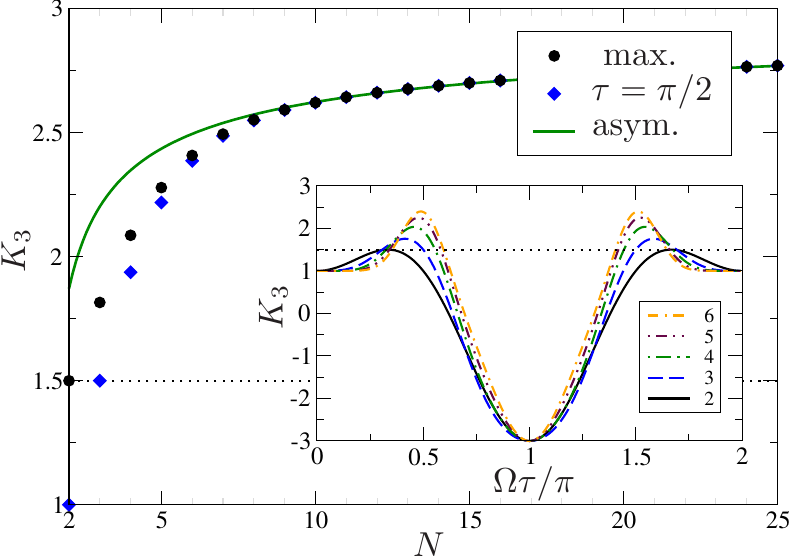}
    \caption{ 
    The Leggett-Garg quantity $K_{3}$ for a spin of length $j=(N-1)/2$ precessing in magnetic field with measurement times $\Omega t_1 = \pi$, $t_2-t_1 = t_3-t_2 =\tau$.
    The measurement is made with $M=N$ projectors (von Neumann scheme) in the the $z$-direction.
    {\bf Inset:}
    $K_3$ as a function of measurement time $\tau$ for various values of $N$.
    For $N=2$, the maximum is familiar qubit or L\"uders bound $K_3^\mathrm{max} =\frac{3}{2}$ (solid line).  For $N=3$, however, the maximum value is 1.7565, and this increases with increasing $N$. 
    {\bf Main panel:}
    The black circles show the maximum value $K_{3}^\mathrm{max}$ as a function of system size $N=2j+1$ for the spin precession model with measurement times as above.
    The blue diamonds show the value of $K_{3}$ with $\tau$ fixed $\Omega \tau = \pi/2$ and the solid line shows the asymptotic behaviour $K_{3}^\mathrm{max} \sim 3 - \sqrt{2/\pi j}$. In the limit $N\to \infty$, $K_{3}^\mathrm{max}$ tends to the algebraic bound of 3.
    \label{FIG:spin}
    }
  \end{center}
\end{figure}

We initialise the system so that at time $t=0$ it is in state  $\ket{\psi(t=0)} = \ket{-j;-j}$ and set the  measurement times as $\Omega t_1=\pi$,  $t_2-t_1=t_3-t_2=\tau$. For $N=2$ we obtain the familiar qubit results \cite{Emary2013}.  For $N=3$, the LGI parameter reads:
\begin{multline}
  K_{3} = 
    \frac{1}{16}+2 \cos\rb{\Omega \tau}
    -\frac{5}{4} \cos\rb{2 \Omega \tau}
    +\frac{3}{16}\cos\rb{4 \Omega \tau}
    ,
\end{multline}
which exhibits the key property in which we are interested --- as  \fig{FIG:spin} shows, this quantity shows a maximum of 
$K^\mathrm{max}_{3} = 1.7565$, clearly in excess of the L\"uders bound.

{\it Asymptotic limit.---} Figure~\ref{FIG:spin} further shows that the maximum value of $K_3$ for this model increases as a function of system size, $N$. In the limit $N \to \infty$, the maximum possible violation is $K_3^\mathrm{max}=3$, as we now show.  With measurement times $\Omega \tau=\frac{1}{2}\pi$,
the correlation functions read \cite{SI}
\beq
  C_{31} 
  &=&
  -1
  ;\quad\quad
  C_{21}
  =
  1 - 2^{1-2j}
  ;
  \nonumber\\
  C_{32} 
  &=&
  1 - 2 \frac{1}{2^{2j}}
  + 4 \frac{1}{2^{4j}}
  -2 \frac{(4j)!}{4^{2j}[(2j)!]^2}
  .
\eeq
The corresponding value of $K_3$ as a function of $N$ is shown in \fig{FIG:spin}.
For finite $N$, this choice of measurement time does not give the maximum violation.  However, this result serves to bound $K_3^\mathrm{max}$ and, for large $j$, the asymptotic behaviour is
\beq
  K_{3} \to 3 - \sqrt{\frac{2}{\pi j}}
  .
\eeq
Thus, at least in the limit that the dimension of the system becomes infinite, the $K_3$ LGI can be violated by quantum mechanics all the way up to the algebraic bound.

{\it Maximum violations.---}Whilst the precessing spin model reveals violations greater than the qubit case can occur, the violations for this system are {\em not} the maximum possible violations at a given $N$ and $M$. Again, this is in contrast with the $M=2$ case where the Rabi oscillation of the qubit provides the maximum violation.

To investigate the true maximum violations as a function of $N$ and $M$, we combine two different methods. The maximum value for a given $M$ can be obtained by means of the maximization method for temporal correlations based on semidefinite programming \cite{Budroni2013}.
This method provides an upper bound valid for any $N$, which is attained for any $N\geq N_{min}$. However, the exact value for $N_{min}$ cannot be extracted from the solution, even though the method provides a state and a set of observables attaining the maximum quantum value \cite{SI}.
 
\begin{figure}[tb]
  \begin{center}
    \includegraphics[width=\columnwidth,clip]{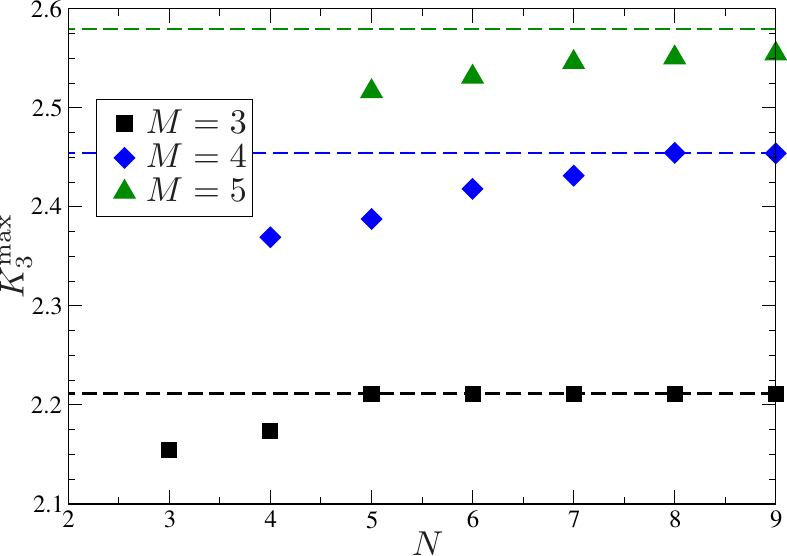}
    \caption{ \label{FIG:maxviolN}
      A plot of the data in Tab.\ref{TAB:data}. The maximum values for each $M$ (from SDP) are shown  as straight lines. 
    }
  \end{center}
\end{figure}

We also pursue a complementary approach in which, for explicit values of $N$ and $M$, we numerically maximise $K_3$ over time-evolution operators $U_{\beta\alpha}$ treated as general $N\times N$ unitary matrices.  The results from these calculations are summarized in Tab.~\ref{TAB:data} and Fig.\ref{FIG:maxviolN}.  We observe that the $M=3$ and $M=4$ bounds from semidefinite programming are saturated at relatively small system sizes, $N=5$ and $N=8$ respectively.

\begin{table}[t]
  \begin{tabular}{|c|c| c |c|c|c| c |c|c|c| c |c|c|c|}
    \hline
    \multicolumn{2}{|c|}{SDP} & & \multicolumn{11}{c|}{MAX}
    \\
    \hline
    $M$ & $K_3^\mathrm{max}$ & &
      $M$ & $N$ & $K_3^\mathrm{max}$ & & $M$ & $N$ & $K_3^\mathrm{max}$ & & $M$ & $N$ & $K_3^\mathrm{max}$\\
    \hline
    2 & $\frac{3}{2}$ && 
      3 &  3 & 2.1547  & &  4 & 4 & 2.3693 && 5 & 5 & 2.5166\\
    3 & 2.211507 && 
      3 & 4 & 2.1736   & &  4 & 5 & 2.3877  && 5 & 6 & 2.5312\\
    4 & 2.454629 && 
      3 & 5 & 2.2115   & &  4 & 6 & 2.4181 & & 5 & 7 & 2.5459\\
    5 & 2.579333 &&
      3 & 6 & 2.2115   & &  4 & 7 & 2.4315 & & 5 & 8 & 2.5506\\
    6 & 2.656005 &&
      3 & 7 & 2.2115   & &  4 & 8 & 2.4545 & & 5 & 9 & 2.5545\\
 
    \hline
  \end{tabular}
    \caption{ 
      The maximum value of the LGI parameter $K_3$  as a function of system size $N$ and number of projectors $M$. The leftmost results  are from the semi-definite programming (SDP) approach, whilst the rest are from direct maximisation (MAX) with fixed $N$ and $M$.
      Here the value assignments $q_m=1-2\delta_{m,-j}$ were used.   In general, the bound changes for different assignments, but except for the case $M=6$, the above choice was found to give the maximum violation.
    \label{TAB:data}
    }
\end{table}


{\it Temporal versus spatial correlations.---} Leggett-Garg inequalities are often referred to as `Bell inequalities in time'; in addition, it is known the L\"uders bound for the $n$-term generalization, for even $n$, of the original Leggett-Garg inequality (\ref{K3intro}) coincides with the Tsirelson bound \cite{Budroni2013} for the corresponding Bell inequalities \cite{Braunstein89,Wehner06}, and noncontextuality inequalities \cite{Araujo13}.  It is therefore a natural question whether the above general measurement scheme can provide excess quantum violation of Bell inequalities.
The answer, however, is negative as can be easily deduced directly from the Tsirelson's proof of the quantum bound \cite{Cirelson1980} or by noticing that the commutativity of the measurements, even when performed sequentially as in contextuality tests, makes irrelevant the post-measurement state and therefore which state-update rule is used.


{\it Discussion.---} We have shown that higher violations of the Leggett-Garg inequality are possible within the framework of standard quantum theory plus projective measurements. This is of fundamental importance since classical theories reproducing, or exceeding, the quantum correlations for temporal scenarios are conceivable and they do not violate any physical principle, as opposed to Bell scenarios where such classical theories involve faster-than-light communication between space-like separated experiments. In fact, in a temporal scenario a classical device with memory, keeping track of the performed measurements and outcomes, can easily saturate the algebraic bound. However, such a device cannot be considered in Leggett-Garg tests since it contradicts the hypothesis of non-invasiveness of the measurement: the memory must be stored on a (possibly auxiliary) physical system. The same argument applies also to the quantum mechanical description of such a device, which is only possible with POVMs \cite{Fritz2010}. Such measurement schemes are, therefore, not meaningful in a Leggett-Garg test.

From an information-theoretic perspective, it is interesting to relate temporal correlations to
the amount of information transmitted through sequential measurements 
\cite{Fritz2010}. While classical devices with memory, and their quantum counterparts based on POVMs, can easily saturate the algebraic bound 
$K_3=3$, the amount of information transmitted trough 
sequential projective measurements, subjected to L\"uders rule, has been proven to obey stricter 
bounds, independent of the system size \cite{Budroni2013}. Our analysis shows that 
degeneracy-breaking projective measurements, as those in von Neumann's scheme, are able to transmit 
more information, which is encoded in the different evolution paths in the set of quantum state, 
and can give rise to perfect correlations (or anticorrelations) in the limit of an infinite 
number of projectors. This is in stark contrast with Bell inequalities, which do not show any 
higher violation when tested with more general type of quantum measurements and are typically 
saturated only in the framework of post-quantum theories \cite{Popescu1994}.

We stress that this analysis does not contradict the conclusions drawn in Ref.\cite{Budroni2013}, information-theoretic principles bounding the temporal correlations for projective measurements may still exist, but such principles must take into account the fact that the bound depends on the number of level accessed via projective measurement.

An application of our results is that of a {\it dimension witness} \cite{Brunner08}: an experimenter can certify that she is able to manipulate at least $M$ levels of a quantum system, if she can violate the bound for $M-1$. Obviously, also the condition of projective measurement must be verified. Notice that our dimension witnesses involve always the same Leggett-Garg inequality and measurement scheme, in contrast to other proposal based on Bell \cite{Brunner08} or noncontextuality \cite{Guehne13} inequalities, and  the prepare-and-measure scenario \cite{Gallego10}, where specific inequalities violated only by high-dimensional systems and involving more complex measurement schemes must be found. 

A further interesting application is the discrimination between L\"uders' and von Neumann's state-update rules \cite{Hegerfeldt2012}, i.e., which one, if any, correctly represents the measurement scenario. A violation of the bound corresponding to $M=2$ shows a contradiction with L\"uders rule. Intermediate cases are possible and can also be investigated with our method.

Moreover, we hope that our results will be a catalyst for experimental investigation of  
higher-dimensional systems and the measurement of  violations of the Leggett-Garg inequalities beyond those achievable either with a single qubit or, indeed, in a Bell scenario.

{\it Acknowledgements.---} We thank J.~Anders, O.~G\"uhne, M.~Kleinmann, and T.~Moroder for discussions. This work has been supported by the EU (Marie Curie CIG 293993/ENFOQI), the BMBF 
(Chist-Era Project QUASAR), the FQXi Fund (Silicon Valley Community Foundation) and the DFG.


\section{Supplemental Information}
\subsection{Asymptotic value of the Leggett-Garg correlator for the precessing spin model}

We here derive the expression for the correlation functions for the spin model with measurement times
$\Omega t_1 = \pi$, $\Omega t_2 = \frac{3}{2}\pi$ and $\Omega t_3 = 2\pi$. 
Defining $R = e^{-i \frac{\pi}{2} J_x}$, the relevant time-evolution operators can be written $U(t_1) = R^2$, $U(t_2) = R^3$, and $U(t_2-t_1) = U(t_3-t_2) = R$.  Starting in state $\ket{-j}$ (we use the shorthand $\ket{m}\equiv\ket{m;j}$ here), the correlation functions read
\beq
  C_{21} 
  &=&
  \sum_{n,m=-j}^j q_n q_m 
  |\ew{m| R |n}|^2 
  |\ew{n| R^2 |-j}|^2
  ;
  \nonumber\\
  C_{31}
  &=&
  \sum_{n,m=-j}^j q_n q_m 
  |\ew{m| R^2 |n}|^2 
  |\ew{n| R^2 |-j}|^2
  ;
  \nonumber\\
  C_{32}
  &=&
  \sum_{n,m=-j}^j q_n q_m 
  |\ew{m| R |n}|^2 
  |\ew{n| R^3 |-j}|^2
  .
\eeq
The matrix $R^2$ has matrix elements such that $R^2\ket{-j} =(-i)^{2j} \ket{+j} $ and $R^2\ket{+j} =(-i)^{2j} \ket{-j}$.  Thus, we obtain
\beq
  C_{21}
  &=&
  \sum_{m=-j}^j q_m 
  |\ew{m| R |j}|^2 ;
  \quad \quad
  C_{31} 
  =
  -1
  \nonumber\\
  C_{32}
  &=&
  \sum_{n,m=-j}^j q_n q_m 
  |\ew{m| R |n}|^2 
  |\ew{n| R^3 |-j}|^2
  .
\eeq
Using the explicit representation of measurement assigments, $q_m = 1-2\delta_{m,-j} $, we can write
\beq
  C_{21}
  &=&
  \rb{\sum_{m=-j}^j
  |\ew{m| R |j}|^2 } - 2 |\ew{-j| R |j}|^2 
  \nonumber\\
  &=& 1 - 2 |\ew{-j| R |j}|^2 
\eeq
The relevant matrix elements are 
\beq
  |\ew{n|R|-j}| =\frac{1}{2^j} \sqrt{\binom{2j}{n+j} }
  ,
\eeq
such that 
\beq
  C_{21}
  &=& 1 - 2^{1-2j}
  .
\eeq
The final term can evaluated as
\beq
  C_{32}
  &=&
  1 - 2|\ew{-j| R^3 |-j}|^2 
  \nonumber\\
  &&
  + 4 |\ew{j| R |-j}|^2 |\ew{-j| R^3 |-j}|^2
  \nonumber\\
  &&
  -2 \sum_n|\ew{-j| R |n}|^2 |\ew{n| R^3 |-j}|^2
  \nonumber
  \\
  &=&
  1 - 2 \frac{1}{2^{2j}}
  + 4 \frac{1}{2^{4j}}
  -2 \frac{(4j)!}{4^{2j}[(2j)!]^2}
\eeq
We have therefore
\beq
  K_{3} 
  &=& 3 - 4^{1-j} + 4^{1-2j} -  \frac{2^{1-4j}(4j)!}{[(2j)!]^2}
\eeq
For large $j$, the latter term can be approximated as $-\sqrt{2/\pi j}$ which then dominates the $j$-dependence.  In the large-spin limit, we have therefore
\beq
  K_{3} \sim 3 - \sqrt{\frac{2}{\pi j}}
  ,
\eeq
which obviously reaches the value $3$ in the $j \to \infty$ limit.

\subsection{Maximization of  temporal quantum correlations via semidefinite programming}

Here we briefly review the optimization method presented in \cite{Budroni2013} for bounding temporal quantum correlations and discuss how to apply it to our case.

Consider a sequence of length $n$ of measurement, with setting denoted as $\mathbf{s}=(s_1,\ldots,s_n)$ and outcomes denoted as $\mathbf{r}=(r_1,\ldots,r_n)$, and ordering such that $r_1,s_1$ label the result and setting for the first measurement, and so on. We denote the corresponding conditional probability as
\begin{equation}\label{eq:pcon}
P(\mathbf{r}|\mathbf{s})=Prob(r_1,\ldots,r_n| s_1,\ldots,s_n)
\end{equation}
For projective quantum measurements, each pair of outcome and setting $(r,s)$ is associated with a projector $\Pi_r^s$, which altogether satisfy $\sum_r \Pi_r^s=\mathbbm{1}$ (completeness) and $\Pi_r^s\Pi_{r'}^s=\delta_{rr'}\Pi_r^s$ (orthogonality). The state-update rule for a measurement with setting $s$ and result $r$ is given by $\rho\rightarrow \Pi_r^s \rho \Pi_r^s / P(r|s)$.

A conditional probability distribution $P(\mathbf{r}|\mathbf{s})$ has a sequential projective quantum representation if there exists a set of operators $\{\Pi_r^s\}$ and a quantum state $\rho$ such that
\begin{equation}\label{eq:condpr}
P(\mathbf{r}|\mathbf{s}) = \trace{\Pi(\mathbf{r}|\mathbf{s}) (\mathbf{r}|\mathbf{s})^\dagger\rho },
\end{equation} 
where $\Pi(\mathbf{r}|\mathbf{s})\equiv \Pi_{r_1}^{s_1}\Pi_{r_2}^{s_2}\ldots\Pi_{r_n}^{s_n}$.

We can now define the matrix of moments as
\begin{equation}\label{eq:matmom}
M_{\mathbf{r}|\mathbf{s};\mathbf{r}'|\mathbf{s}'}\equiv \ew{ \Pi(\mathbf{r}|\mathbf{s}) (\mathbf{r'}|\mathbf{s'})^\dagger}.
\end{equation}
$M$ is positive semidefinite, i.e. it has no negative eigeinvalues, denoted as $M\succeq 0$, and it satisfies linear relations of the form $M_{{\bf r|s; k|l}} = M_{{\bf r^\prime|s^\prime; k^\prime|l^\prime}}$ if the corresponding entries are equal as a consequence of the completeness and orthogonality properties of the corresponding projectors. Notice that diagonal elements of $M$ corresponds to conditional probabilities for sequential projective measurement as in eq. (\ref{eq:condpr}). 

We can now state our optimization problem as a semidefinite program (SDP)
\begin{eqnarray}\label{eq:sdp}
\mbox{maximize:}\ && \sum_{ij} c_{ij} M_{ij},
\\ \nonumber 
\mbox{s.t. :}\ && M=M^T \succeq 0 \mbox{ and } 
\sum_{ij} F_{ij}^{(k)}M_{ij}=g_k,
\end{eqnarray}
where the coefficients $\{F_{ij}^{(k)},g_k\}$ give the linear constraint discussed above, and 
$\{c_{ij}\}$ define the linear function of the matrix entries to be maximized, which will 
correspond to the products $q_lq_m$ in Eq. (\ref{QQ}) and the coefficients in the Leggett-Garg inequality Eq. (\ref{K3intro}).

The maximization is taken over all the semidefinite matrices satisfying a set of linear conditions, 
but it can be proven that from any solution of the SDP a quantum state and a set of observables 
attaining the same value for the linear function can be constructed. The bound obtained is 
therefore tight. Moreover, the bound is valid for any dimension $N$ of the Hilbert space, and it is 
attained for any $N\leq N_{min}$. However, this procedure gives a state and a set of observables 
defined in a Hilbert space of dimension equal to the rank of the solution matrix $M$, which, in 
general, is bigger than $N_{min}$.

In general, the above method gives bounds more general than those involved in Leggett-Garg tests. In fact, Leggett-Garg tests involve the measurement of the same observable at different time, i.e. the different operators must be connected via unitary transformations. However, since the bound is independent of the dimension and we are considering observables with the same spectrum, we can always extend the Hilbert space such that each eigenvalue has the same degeneracy for every observables. Then, clearly, there exist unitary transformations connecting the different observables (a unitary operator, by definition, transform orthonormal basis in orthonormal basis). In general, such unitary operators will not commute, therefore, the corresponding time-evolutions cannot be given by a time-independent Hamiltonian $H$ via the operator $e^{-iHt}$, but a more general time-evolution must be considered.

\end{document}